\documentclass[epj,english,twocolumn]{webofc}
\usepackage[varg]{txfonts}   
\usepackage{hyperref}        
\usepackage{multirow}                            
\usepackage{graphicx}                            
\usepackage{latexsym}                            
\usepackage{amsfonts}                            
\usepackage{amssymb}                             
\usepackage{amsmath}                             
\usepackage[mathscr]{eucal}                      
\usepackage{latexsym}                            
\usepackage[mathscr]{eucal}                      

\newcommand{\bc}{\begin{center}}
\newcommand{\ec}{\end{center}}
\newcommand{\be}{\begin{equation}}
\newcommand{\ee}{\end{equation}}
\newcommand{\bea}{\begin{eqnarray}}
\newcommand{\eea}{\end{eqnarray}}

\woctitle{Heavy Ion Accelerator Symposium 2013}
\begin{document}
\title{QCD and a new paradigm for nuclear structure}
\author{A.~W.~Thomas\inst{1}\fnsep\thanks{Supported by the Australian Research Council through grants DP150103101 and CE110001104.} 
}

\institute{CSSM and ARC Centre of Excellence for Particle Physics at the
Terascale,\\ School of Chemistry and Physics,
University of Adelaide,
Adelaide SA 5005, Australia}

\abstract{%
\par
We review the reasons why one might choose to seriously re-examine 
the traditional approach to nuclear theory where nucleons are treated 
as immutable. This examination leads us to 
argue that the modification of the structure
of the nucleon when immersed in a nuclear medium is fundamental to how
atomic nuclei are built. Consistent with this approach we suggest key 
experiments which should tell us unambiguously whether there is such 
a change in the structure of a bound nucleon. We also briefly report on
extremely promising recent calculations of the structure of nuclei 
across the periodic table based upon this idea.
}
\maketitle
\section{Introduction}
\label{intro}
Since the discovery of the neutron in the 1930s, the overwhelming 
majority of theoretical studies of nuclear structure have adopted 
the hypothesis that the protons and neutrons inside a nucleus are 
immutable objects whose internal structure never changes. These 
immutable objects interact through non-relativistic two- and 
three-body forces and the challenge is primarily to accurately 
solve the many-body problem. The phenomenological forces used 
include physics such as Yukawa's pion exchange and as a consequence 
the precise calculation of observables may require the inclusion 
of exchange current corrections. 

Beginning with the famous one-boson-exchange 
potentials~\cite{Ueda:1969er}, it became clear 
that the dominant part of the intermediate range attraction between 
nucleons had a Lorentz scalar, isoscalar character, which was 
phenomenologically represented by the exchange of a $\sigma$ meson. For 
decades this meson was viewed as an artifact involving an unphysical 
meson used purely for convenience. However, careful dispersion relation 
treatments of $\pi N$ scattering in the past decade have shown that 
this state does indeed exist~\cite{Ananthanarayan:2000ht}. 
Confirmation of this Lorentz scalar, isoscalar 
character of the intermediate range attraction in the $NN$ force also 
came from dispersion relation studies by groups 
in Paris~\cite{Cottingham:1973wt}, 
Stony Brook and elsewhere.
Walecka and co-workers exploited the Lorentz scalar nature of 
the $NN$ attraction and the Lorentz vector character of the short 
range repulsion to build a very successful, 
{}fully relativistic theory of nuclear matter~\cite{Walecka:1974qa} 
and later finite nuclei~\cite{Serot:1979cc}. 
Here too the nucleons were immutable.

All this was very satisfactory but for one vexatious issue. At nuclear 
matter densities the typical mean scalar field strength felt by a 
bound nucleon in the Walecka model is of order 500 MeV. {\em This is a huge
number.} As a consequence, the effective mass of the bound nucleon is  
only one half of its free mass. 

At around the same time as Walecka and collaborators developed their model, 
the theory of the strong interaction underwent a revolution. Quantum 
Chromodynamics was developed as a local gauge field theory built on 
color. It became clear that, by analogy with Rutherford's work on the 
nucleus within the atom, the natural 
explanation of the discovery of scaling at 
SLAC in the late 60's~\cite{Bloom:1969kc}
was that the nucleon too was primarily empty 
space containing point-like quarks.

{\em From this more fundamental point of view 
the huge scalar field experienced 
by a bound nucleon is even more challenging. How can it be that the 
exchange of a scalar meson, which must couple to the confined quarks 
in the nucleon with such strength, can have no effect on the internal 
structure of the nucleon, which after all is far from point-like?}

Considerations like these led Guichon~\cite{Guichon:1987jp} 
to propose a dramatically different 
approach to nuclear binding, the Quark Meson Coupling (QMC) model, 
where the effect of the mean scalar field 
generated by other nucleons is treated self-consistently in solving 
for the wave function of each confined quark.  
Taking the simplest form for the coupling of the $\sigma$ and $\omega$ 
mesons to quarks confined in the MIT bag 
model~\cite{Chodos:1974je,Thomas:1982kv},
means that in nuclear matter the vector field simply 
shifts the definition of the energy, while the scalar field modifies 
the Dirac wave function. {\em This difference in the effect of the two 
Lorentz components of the nuclear mean field is crucial, as their 
effects more or less cancel when it comes to the total energy but for the 
quark motion (or loosely speaking, wave function) the scalar field is not 
cancelled.} 

A critical effect of the change in the quark wave function induced by 
an attractive scalar field is that the size of the lower  
Dirac component increases. In turn this reduces the value of 
$\int dV \bar{\psi} \psi$, which defines the overall  strength with which 
the scalar field couples to the nucleon. 
This process is completely analogous to the way an atom rearranges its 
internal structure to oppose an applied electric field. Thus the 
parameter calculated within any particular quark model which describes 
this is called the ''scalar polarizability'', $d$. The overall scalar 
coupling to the nucleon is written in the simplest approximation as
\begin{equation}
g_{\sigma N}(\sigma) = g_{\sigma N}(0) - \frac{d}{2} 
(g_{\sigma N}(0) \sigma)^2 \, .
\label{eq:scalarpol}
\end {equation}
In the MIT bag model $d \approx 0.22R$, with $R$ the bag radius.

This behaviour is very straightforward and appears in all relativistic
quark models used so far. Nevertheless, in terms of nuclear structure it 
is profound. Whereas the repulsion felt by each nucleon grows linearly 
with density, the scalar attraction saturates as the density rises and 
one naturally finds saturation of nuclear matter. This mechanism is both 
new and extremely effective. As a result the mean scalar field felt 
by a nucleon at the saturation density of nuclear matter is just a few 
hundred MeV, much lower than that found in the Walecka model.

Philosophically, this approach is radically different from anything done 
before because the colourless clusters of quarks which occupy single 
particle levels in nuclear matter may have nucleon quantum numbers 
but their internal structure is modified. Almost immediately it was 
shown~\cite{Thomas:1989vt} that this change could account for 
the key features of the famous
nuclear EMC effect, discovered in the early 80's.

Later the model was developed further by Guichon, Rodionov 
and Thomas~\cite{Guichon:1995ue} to 
correctly treat the effect of spurious 
centre of mass motion in the bag, which 
had resulted in anomalously small $\omega N$ couplings. In
the same paper the model was also extended 
to finite nuclei, showing very naturally how one obtains realistic 
spin orbit forces. 
{}Finally, since the model is built at the quark level, using the same 
quark model, with the same quark-meson coupling constants, one can derive 
the properties of {\em any} bound hadron. For example,
it shows very naturally         
why the spin-orbit force for the $\Lambda$ hyperon 
is extremely small~\cite{Tsushima:1997cu,Tsushima:1997rd}.
{}For a complete overview of the phenomenological consequences of the 
QMC model we refer to the review by 
Saito {\em et al.}~\cite{Saito:2005rv}.

With the motivation for the QMC approach clearly established, one 
is naturally led to the following lines of investigation. First, 
given the success of the conventional approach to nuclear structure 
based upon non-relativistic two- and three-body forces, it is natural 
to ask how that is related to QMC. We address this in section 2. Second, 
one may also ask what evidence there is to support the at first sight  
radical idea that the clusters of quarks bound in shell model orbits 
actually have internal structure different from that of a free nucleon.
This is addressed in Section 3, where we anticipate the results 
of a critical experiment performed at Jefferson Lab, which are expected to 
appear soon. Section 4 summarises this new approach to the structure 
of the atomic nucleus and looks to further consequences of it. 

\section{Nuclear structure: a new force of the Skyrme type}
It is worthwhile to begin with some remarks on the application of 
effective field theory (EFT) to nuclear structure, since that also is 
often treated as containing all of the 
consequences of QCD~\cite{Weinberg:1991um}. Certainly 
the systematic application of chiral effective field theory to the $NN$ 
and $NNN$ forces and hence to nuclear structure has proven quite 
powerful. Such an approach is built upon the symmetries of QCD and 
is often considered to be equivalent to it. The problem is that the 
EFT approach needs some power counting scheme, which is a purely 
human construction. It also needs a set of 
hadronic degrees of freedom (dof) 
and that choice too is at the whim of the user. Finally, the EFT typically 
applied to nuclear problems is non-relativistic.

The usual choice  of dof are nucleons and pions. If these are 
indeed the appropriate dof one is in luck. However, given the remarks 
in the Introduction, where we saw that on model independent grounds 
the intermediate range attraction between nucleons is a rather large 
Lorentz scalar, this is not so obvious. The attractive scalar and repulsive 
vector forces may cancel (in the central component of the nuclear force) 
to produce a relatively small amount of binding but
the effect of those two components on the internal structure of a 
nucleon is completely different. 

In an EFT the 
only way to include the effect of a change in the structure 
of a bound nucleon at the 
level of QCD is to include nucleon excited states amongst the dof. 
Typically this is limited to the $\Delta$ resonance, where we do 
know the relevant couplings quite well. However,
given that the $\sigma$ meson has quantum numbers $0^{++}$, 
one may expect that the inclusion of excitations like the Roper 
resonance~\cite{Leinweber:2015kyz}
may be relevant. Unfortunately,
we have so little knowledge of that 
state that, at the present time,
it would be very difficult to include it in an EFT framework 
in any reliable manner.  

As a consequence, building an EFT of nuclei based upon nucleon and pion 
dof may not be as accurate an expression of QCD as 
it may appear at first sight.

An alternative approach to developing an EFT for nuclear structure 
is based on the density functional approach. There one starts with the 
QMC model itself and develops a density functional equivalent to it.
From this one can use the machinery developed around the Skyrme 
forces~\cite{Vautherin:1971aw}, 
which have proven so successful in the study of both nuclear 
structure and reactions.

Indeed, using the density functional approach it has proven possible to 
develop a clear connection between the self-consistent
treatment of in-medium hadron structure and the existence of
many-body~\cite{Guichon:2004xg} or density
dependent~\cite{Guichon:2006er} effective forces.  Dutra {\it et
  al.}~\cite{Dutra:2012mb} critically examined a variety of
phenomenological Skyrme models of the effective density dependent
nuclear force against the most up-to-date empirical constraints. 
Amongst the few percent of the Skyrme forces studied which satisfied 
all of these constraints, the Skyrme model SQMC700, 
was unique in that it was actually {\em derived
from the QMC model} and hence incorporated the effects of
the internal structure of the nucleon and its modification in-medium.

Very recently, Stone, Guichon, Reinhard and  
Thomas~\cite{Stone:2016qmi} 
carried out a systematic study of the properties of atomic nuclei 
across the whole periodic table using the 
new, effective, density-dependent $NN$ force 
derived from the QMC model~\cite{Guichon:2006er}. 
The study began by defining those combinations of the three fundamental 
couplings in the model (namely the $\sigma, \omega$ and $\rho$ couplings to 
the up and down quarks) which reproduce the saturation density, binding 
energy per nucleon and symmetry energy of nuclear matter within the 
empirical uncertainties on these quantities. Then, a search was carried out 
for the set of three parameters satisfying this nuclear matter constraint 
which best described the ground-state properties of a selection of more
than 100 nuclei across the  entire periodic table.

The root mean-square deviation of the fit from the actual binding 
energy for this set of nuclei was just 0.35\%. For the superheavy 
nuclei where the binding energies are known, the deviation was a 
mere 0.1\%. This level of agreement with the empirical binding 
energies is remarkable, in that it is 
comparable with the very best phenomenological 
Skyrme forces which have typically 11 or more adjustable parameters.

Not only does this derived effective $NN$ force satisfactorily describe 
binding energies but going beyond the nuclei used in the fit it 
accurately describes the evolution of quadrupole deformation across 
isotopic chains, including shell closures. It also proved capable 
of describing the observed shape co-existence of prolate, oblate 
and spherical shapes in the Zr region. Finally, it naturally gave 
a double quadrupole-octupole phase transition in the Ra-Th region.

These are remarkable successes given the extremely small number of 
parameters and this suggests that it would be worthwhile to apply 
this derived effective force across a variety of challenges in 
modern nuclear physics.

\section{Experimental tests}
Almost immediately after the creation of the QMC model it was 
applied~\cite{Thomas:1989vt} to the modification of 
the valence quark distribution 
in nuclei discovered by the European Muon 
Collaboration (EMC), known as the 
EMC effect~\cite{Aubert:1983xm}. 
That early work was based on the MIT bag model, for 
which the calculation of structure functions 
is possible within some approximations~\cite{Signal:1989yc} 
but complicated. More
recently, the generalization of the QMC model to the NJL model, suggested 
by Bentz and Thomas~\cite{Bentz:2001vc}, 
has also been applied  to the EMC effect with similar 
success~\cite{Cloet:2006bq}. 
The modification of the quark wave functions within the bound 
nucleons, because of the applied mean scalar field, naturally suppresses 
the valence distributions at large Bjorken $x$.

While this approach is the {\em only} quantitative model of nuclear 
structure which is able to describe the nuclear EMC effect, it is 
not yet universally accepted as the explanation for it. For example, 
it has recently been suggested that the 
entire EMC effect should be attributed 
to an as yet uncalculable modification of the nucleons involved in 
short-range correlations~\cite{Wang:2014bba}, 
while the rest of the nucleons  
apparently remain totally unchanged.

Another feature of this approach to nuclear structure is that the elastic 
form factors of the nucleon are also modified in-medium~\cite{Lu:1998tn}.
Using the QMC model, predictions were made almost 20 years ago for the
experiment being planned at Jefferson Lab to measure the ratio of 
the electric to magnetic form factors 
of a proton bound in $^4$He~\cite{Lu:1997mu}.
A decade later the measurements were in remarkably good agreement with 
those predictions~\cite{Strauch:2002wu,Paolone:2010qc,Udias:2000ig}, 
showing a significant medium modification.
However, after the data appeared it was shown that it could also be fit 
by adding an unusually large polarised charge exchange correction.  
Although we are aware of no data supporting that proposed correction 
and no proposal to check it experimentally, it has muddied the waters 
sufficiently that this cannot yet be regarded as a ''smoking gun''.

\begin{figure}[!tb]
\centering
\includegraphics[width=8cm]{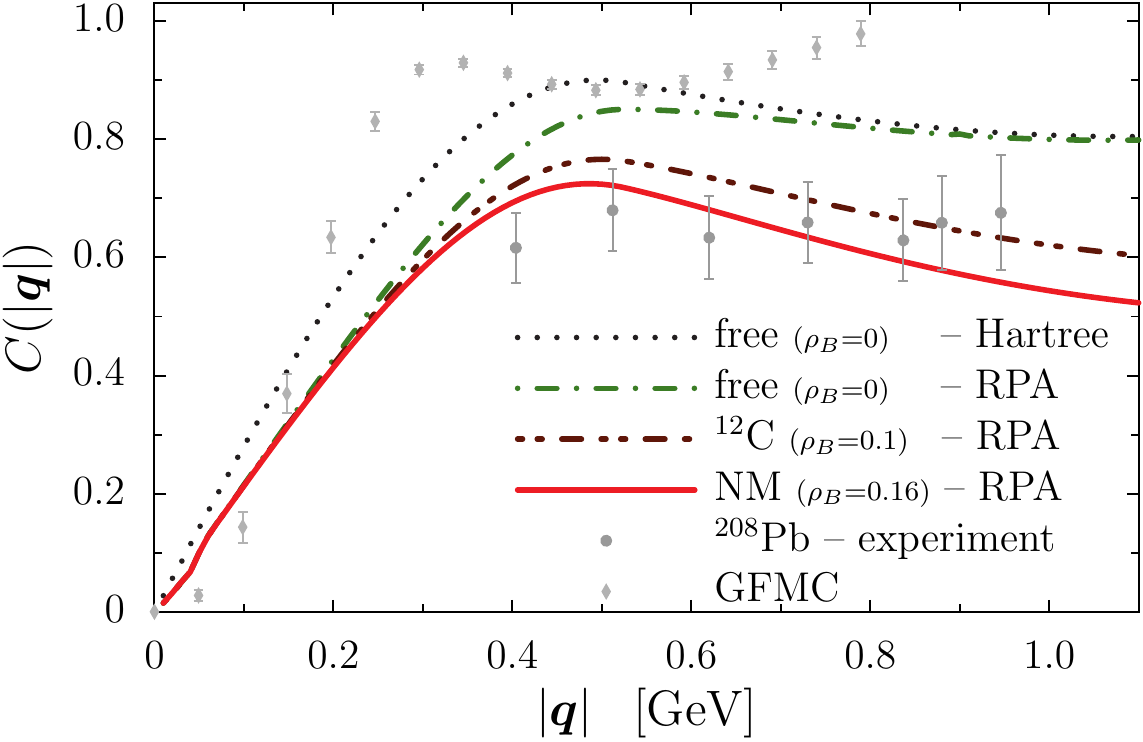}
\caption{Predictions 
(from Ref.~\cite{Cloet:2015tha}) for 
the Coulomb sum rule as a function of three momentum 
transfer for nuclear matter at densities corresponding to $^{12}$C and 
$^{208}$Pb, with or without the effect of the in-medium modification 
of the nucleon electric form factors. 
Also shown are the GFMC 
calculations for $^{12}$C 
(small points \cite{Lovato:2013cua}) 
and older experimental data for 
$^{208}$Pb~\cite{Morgenstern:2001jt,Zghiche:1993xg}.}
\label{fig:CSR}       
\end{figure}

Another suggestion, which seems far less susceptible to unknown 
nuclear corrections, involves the measurement of the longitudinal 
response function measured in inelastic electron 
scattering~\cite{Morgenstern:2001jt}. That 
was also examined in the late 90's on the basis of the modification 
of the electric form factor of the proton~\cite{Saito:1999bv}, 
already mentioned. Very 
recently, inspired by the proposal of Meziani 
and collaborators~\cite{JLab-expt} to 
make a definitive measurement of this quantity for several nuclei 
across the periodic table, this response function and the associated 
Coulomb sum rule of McVoy and van Hove~\cite{McVoy:1962zz} 
were investigated using the NJL 
model to describe the structure of both 
the free and bound nucleons~\cite{Cloet:2015tha}. 
This work not only treated self-consistently 
the modification of bound nucleon structure 
resulting from the mean scalar field but it also included 
a state-of-the-art treatment of relativistic 
corrections and RPA correlations. The results are illustrated 
in Fig.~\ref{fig:CSR}.

At high values of the momentum transfer the effect of relativity and of 
the medium modification of the electric form factor of the proton 
in particular are both very significant. The older data certainly favours 
the new calculations and it is clearly vital to have the results of 
the comprehensive new experiment from Jefferson Lab as soon as possible. 
The beauty of this particular measurement is that it appears to be extremely 
insensitive to other nuclear corrections, including the effect of 
short-range correlations.

\section{Summary}\label{sec-3-Summary}
We have presented a compelling argument that within the framework 
of QCD one is naturally led to the conclusion that the structure of a 
bound nucleon must differ from that in free space. 

This idea has been used to derive, starting from the quark level, 
a new, density-dependent effective nuclear force which has proven 
remarkably accurate in describing the properties of finite nuclei 
across the entire periodic table, while at the same time reproducing 
the known properties of nuclear matter. 
We trust that these remarkable results will inspire 
a great deal more work on nuclear structure within this framework 
over the coming years.

We have seen that within the quantitative models of nuclear structure 
that have been developed within this approach, using either the MIT bag 
or the NJL model to describe nucleon structure, one finds a natural 
explanation of the nuclear EMC effect. There are also predictions for the 
modification of the electromagnetic form factors of the bound nucleon, for 
which the most unambiguous test is the Coulomb sum rule. 
There is an expectation that definitive new data for this will come from 
Jefferson Lab in the near future.
 
Finally, we briefly mention a number of other consequences of this 
approach to nuclear structure which are both fascinating and the subject  
of experimental investigation in the near future. For example, a careful study 
of nuclear structure functions has shown that this approach
predicts an important isovector 
component of the nuclear EMC effect~\cite{Cloet:2009qs}. For a 
nucleus like $^{56}$Fe this 
leads to a correction to the Paschos-Wolfenstein relation which is of 
the sign and magnitude to reduce the NuTeV anomaly by more than 
one standard deviation. These predictions will be tested directly in 
future measurements of parity violation~\cite{Cloet:2012td} 
at Jefferson Lab following 
the 12 GeV upgrade.

Within this approach one also finds a remarkably large nuclear modification 
of the spin dependent parton distributions of 
the nucleon~\cite{Cloet:2005rt}.
Again, future 
experiments planned at Jefferson Lab will test this through the measurement 
of the spin structure functions of light nuclei with an unpaired proton.

In conclusion, we stress that while one can derive effective $NN$ forces 
which can be used in traditional nuclear structure calculations, {\em the 
underlying physics constitutes a new paradigm for nuclear theory}. The quark 
clusters which occupy shell model orbits in finite nuclei  have internal 
structure which depends on the local scalar field -- they are 
{\em not immutable}. This simple observation, which is entirely natural 
within the framework of QCD, explains the saturation of nuclear matter
and the nuclear EMC effect and predicts 
a dramatic reduction in the Coulomb sum rule as well as a multitude of other 
phenomena which will be subject to experimental study in the coming decade.

\section*{Acknowledgements}
I am indebted to the many collaborators who have contributed to the  
understanding of this approach to nuclear structure, particularly 
P.~A.~M.~Guichon, W.~Bentz, I.~Clo\"et, K.~Saito, J.~Stone and 
K.~Tsushima. This work was supported by the University of Adelaide 
and by the Australian Research Council through the ARC Centre of 
Excellence for Particle Physics at the Terascale (CE110001104), 
an ARC Australian Laureate Fellowship (FL0992247) and DP150103101.


\end{document}